\documentclass[a4paper,prd, 11pt,amsfonts,amssymb,amsmath,nobibnotes,nofootinbib]{revtex4}

\pdfoutput=1

\usepackage[a4paper,scale=0.73]{geometry}
\usepackage{graphicx}

\newcommand{\glin}{g}
\newcommand{\longpage}[1][1]{}
\newcommand{\shortpage}[1][1]{}
\newcommand{\fud}{\frac{1}{2}}
\newcommand{\D}{{\mathcal{D}}}

\DeclareMathOperator{\Tr}{Tr}
\DeclareMathOperator{\sign}{sign}
\DeclareMathOperator{\PV}{P}

\newcommand{\expp}[1]{ \mathop\mathit{e}\nolimits^{#1}}
\newcommand{\dert}[3][]{\frac{\mathrm{d}^{#1} #2}{\mathrm{d} #3^{#1}}}
\newcommand{\derp}[3][]{\frac{\partial^{#1} #2}{\partial #3^{#1}}}
\newcommand{\derpp}[3]{\frac{\partial^2 #1}{\partial #2 \, \partial #3}}

\newcommand{\av}[1]{\langle #1 \rangle}

\newcommand{\ud}[2][]{\textrm{d}^{#1}{#2}\,}

\newcommand{\vd}[2][]{\textrm{d}^{#1}{#2}}
\newcommand{\Eqref}[1]{Eq.~\eqref{#1}}
\newcommand{\ie}{\emph{i.e.}}
\newcommand{\eg}{\emph{e.g.}}
\renewcommand{\Re}{\mathop\mathrm{Re}\nolimits}
\renewcommand{\Im}{\mathop\mathrm{Im}\nolimits}
\newcommand{\vect}{\mathbf}

\newcommand{\GF}{G_\mathrm{F}}

\newcommand{\SigmaR}{\Sigma_\mathrm R}

\newcommand{\Gret}{{G}_\mathrm R}

\newcommand{\GR}{\Gret}

\newcommand{\SigmaN}{\Sigma^{(1)}}
\newcommand{\ti}{{t_\text{i}}}
\newcommand{\tf}{{t_\text{f}}}

\newcommand{\GA}{G_\text{A}}

\providecommand{\citep}{\cite}

\begin{document}
\title{Quantum Brownian representation for the quantum field modes}
\author{Daniel Arteaga}
\email{darteaga@ub.edu}
\altaffiliation{also at Barcelona Media -- Centre d'Innovaci\'o.}
\affiliation{Departament de F\'\i sica Fonamental and Institut de Ci\`encies del Cosmos, Facultat de F\'\i sica, Universitat de  
Barcelona, Av.~Diagonal 647, 08028 Barcelona (Spain)}

\begin{abstract}
	When analyzing the particle-like  excitations in quantum field theory  
it is natural to regard the field mode corresponding to the particle  
momentum as an open quantum system, together with the opposite  
momentum mode.
	Provided that the state of the field is stationary, homogeneous and  
isotropic,   this scalar two-mode system can be equivalently  
represented in terms of a pair of quantum Brownian oscillators under a  
Gaussian approximation.  In other words, the two-mode system behaves  
as if it were  interacting linearly with some effective environment.  
In this paper we build the details of the effective linear coupling  
and the  effective environment, and argue that this quantum Brownian  
representation provides a simple, universal and non-perturbative  
characterization of any single particle-like excitation. As immediate applications of the equivalence, we reanalyse the interpretation of  
the  self-energy in terms of decay rates in a general background state, and present the master equation for the field mode corresponding to the particle momentum.
\end{abstract}

\maketitle
\section{Introduction and motivation } \label{sect:OQS}

Quantum fields  can be regarded from the viewpoint of open quantum  
systems \cite{Davies,BreuerPetruccione,GardinerZoller,Weiss}. The  
degrees of freedom of the field which are relevant for the physical  
problem in question constitute the reduced subsystem and the rest form the  
environment.
If there are several fields in interaction, and the object of interest  
is a particular field, it is natural to trace over the other  
environment fields. For instance, in electrodynamics one can study the  
so-called Euler-Heisenberg effective action for the photons  
\cite{ItzyksonZuber}, considering  the electrons as the environment,  
or the complementary case, in which the electrons are taken as the  
system of interest and the photons are integrated out  
\cite 
{WheelerFeynman49,CaldeiraBarone91,Anastopoulos97,AnastopoulosZoupas97}. Similarly, in  
stochastic gravity
\cite{CalzettaHu94,HuMatacz95,HuSinha95,CamposVerdaguer96,CalzettaEtAl97,LombardoMazzitelli97,MartinVerdaguer99a,MartinVerdaguer99c,MartinVerdaguer00,CalzettaEtAl01,HuVerdaguer03,HuVerdaguer04} the  
system of interest is the gravitational field and the matter fields  
are integrated out. In many other circumstances it is natural to  
consider as the reduced system the modes of the quantum field which  
are below some ultraviolet cutoff, with the ultraviolet modes  
constituting the environment. This approach  has been used, for  
instance, studying bubble nucleation  
\cite{CalzettaRouraVerdaguer01,CalzettaRouraVerdaguer02}, analyzing  
decoherence in field theory  
\cite{LombardoMazzitelli96,GreinerMuller97,LombardoEtAl03} and in  
inflationary cosmology  
\cite 
{Matacz97,TanakaSakagami97,LombardoLopezNacir05,ZanellaCalzetta06}. As  
these references illustrate, the open quantum system point of view has  
often provided new tools and insights to different field theory  
problems, specially when dealing with states different than the  
Minkowski vacuum.

The propagation of a particle-like excitation over a given field  
background is another situation in which there is a natural system--environment separation: the field mode corresponding to the particle  
momentum is the object of interest, and the rest of modes of the  
field, along with any other field in interaction, form the environment.
In the Minkowski vacuum, the K\"all\'en-Lehmann representation of the  
propagator \cite{Peskin,ItzyksonZuber,BrownQFT,WeinbergQFT} provides a  
complete description of single particle excitations from the field  
theory perspective, but there was no equivalent analysis in thermal  
backgrounds  
\cite{LatorreEtAl95,ArteagaEtAl04a,ArteagaEtAl05,LeBellac,BrosEtAl07},  
or in curved spacetimes  
\cite{DrummondHathrell80,Shore03b,Arteaga07a,ArteagaEtAl07a}. The open  
quantum system approach proves a useful tool when reanalising the  
propagation of particle-like excitations in non-vacuum or non-
Minkowski backgrounds \cite{Arteaga08a,Arteaga08b}.

Therefore, in this paper we elaborate on the open quantum description of the  
field mode corresponding to the particle momentum, with the goal of developing useful techniques for the analysis of particle and quasiparticle excitations. We concentrate on the basic methods and tools, and give example applications; the general analysis of the particle-like  
excitations in non-vacuum or non-flat backgrounds is left for separate  
publications \cite{Arteaga08a,Arteaga08b}.

We would like to consider the field  mode as the system and the  
remaining modes as the environment. However, it proves difficult to   
implement this system--environment division directly, since the mode-decomposed field $\phi_\vect p$ is not real, but is a complex quantity  
obeying the contraint $\phi_\vect p = \phi^*_{-\vect p}$ and acting on  
both the Hilbert space sector with momentum $\vect p$ and the sector  
with momentum $-\vect p$. Instead of focusing on a single mode, given  
that the field naturally links modes with opposite momentum, we shall  
choose as the system of interest any two modes with a given opposite  
momentum, and as the environment the remaining modes of the field, as  
well as the modes of any other field in interaction. Namely, given a  
single scalar field $\phi$, the system degrees of freedom are the two  
modes $\phi_\vect p$ and $\phi_{-\vect p}$, and the other modes $\phi_ 
\vect q$, with $\vect q \neq \pm \vect p$, form the environment. A  
similar system--environment division has been used in Refs.~\cite{CampoParentani05a,CampoParentani05b,CampoParentani04} in an  
inflationary context.

The main goal of this paper is to show that this two-mode system  
behaves, under certain assumptions, as if it was an open quantum  
system interacting linearly with some effective environment, even if  
the original field interaction is non-linear.
The paradigm of linear open quantum system is the quantum Brownian  
motion (QBM) model,
whose system of interest is a non-relativistic particle interacting  
linearly with an infinite bath of
harmonic oscillators. This model has had many applications in  
different contexts, among which one may mention the quantum to  
classical transition \cite{UnruhZurek89,RomeroPaz97}, the escape from  
a potential well  
\cite 
{CaldeiraLeggett81,CaldeiraLeggett83a,ArteagaEtAl03,CalzettaVerdaguer06}, the Unruh  
effect \cite{MassarParentaniBrout93,HuMatacz94} or quantum optics  
\cite{WallsMilburn,GardinerZoller}. In an influential paper Caldeira  
and Leggett
\cite{CaldeiraLeggett83b} applied the influence functional model
of Feynman and Vernon \cite{FeynmanVernon63,FeynmanQMPI} to the QBM  
model. The QBM model can be generalised to encompass the general class  
of linearly interacting open quantum systems  
\cite{HuPazZhang92,HuPazZhang93}.

The results of this paper can bee seen as an explicit implementation  
of the statement by Hu and Matacz \cite{HuMatacz94} that the motion of  
a Brownian particle can be used to depict the behaviour of a single  
quantum field mode.

We will allow for stationary and isotropic, but otherwise arbitrary,  
states for the field (a Gaussian approximation will be also assumed  
later on). Field theory with arbitrary field states can be studied  
within the closed time path (CTP), or \emph{in-in}, method, which was  
originally proposed by Schwinger \cite{Schwinger61} and Keldysh  
\cite{Keldysh65}. The most characteristic feature of the CTP method,  
in contrast to the conventional \emph{in-out} method, is the doubling  
of the number of degrees of freedom.

The paper is organised as follows. In Sect.~\ref{sect:QFTasOQS} the  
relevant system--environment separation is discussed. In Sect.~\ref{sect:QBManalogy}, 
the central section of this paper, we present  
and analyse the QBM representation for the field modes, given the  
structure of the generating functional and the two-point propagators.  
In Sect.~\ref{sect:application} we illustrate the utility of the QBM representation with two example applications. First, we rederive the interpretation of the  
imaginary part of the self-energy. Second, we build and analyze the master equation for the field modes corresponding to the particle momentum. The main body of the paper ends with Sect.~\ref{sect:discussion}, where we summarise and discuss the main  results. In order not to break the continuity of the exposition, some  
details of the analysis are left for the appendices, which also  
provide background reference material which fixes the notation and   
makes the paper relatively self-contained. In appendix \ref{app:CTP}  
we briefly present the CTP approach to field theory and apply it to  
the analysis of the structure of the two-point propagators, and in  
appendix \ref{app:QBM} we introduce the theory of linear open quantum  
systems, focusing also in the structure of the two-point propagators.  
The analysis of the propagators done in the appendices is important  
for the discussion in Sect.~\ref{sect:QBManalogy}.

Throughout the paper we work with a system of natural units with $ 
\hbar =c =1$, denote quantum mechanical operators with a hat, and use  
a volume-dependent normalisation in the definition of the field modes  
[see \Eqref{ModeDecomp} below]. The same symbol will be used for a  
quantity and its Fourier transform whenever there is no danger of  
confusion.

\section{Field modes regarded as open quantum systems}  
\label{sect:QFTasOQS}

Let us now present the relevant system--environment separation. For  
concreteness, we consider a self-interacting field theory model  
consisting of a single scalar field $\phi$, although results can be  
straightforwardly extended to any number of fields. The field $\phi$  
can be decomposed in modes according to
\begin{equation}\label{ModeDecomp}
	\phi_\vect p = \frac{1}{\sqrt{V}} \int \ud[3]{\vect x}  \expp{-i  
\vect p \cdot \vect x} \phi(\vect x),
\end{equation}
where $V$ is the volume of the space, a formally infinite quantity.
The factor $V^{-1/2}$ in the definition of $\phi_\vect p$ is chosen so  
that the propagators verify:
\begin{equation*}
	G_+(t,t';\vect p) =  \int \ud[3]{\vect x} \expp{-i\vect p \cdot \vect  
x} \langle 0 | \hat\phi(t,\vect x) \hat\phi(t',\vect 0) |0\rangle =  
\av{ 0 |\hat\phi_\vect p (t) \hat\phi_{-\vect p}(t') |0}.
\end{equation*}

As stated in the introduction, given a particular momentum $\vect p\neq \vect 0$,  
the system is composed by the two modes $\phi_\vect p$ and   $\phi_{- 
\vect p}$, and the environment is  
composed by the other modes of the field, $\phi_\vect q$, with $\vect  
q \neq \pm \vect p$. Should there be other fields in interaction of  
any arbitrary spin, the modes of these additional fields would also  
form part of the environment.

The Hilbert space can be decomposed as $\mathcal H = \mathcal H_ 
\text{sys} \otimes \mathcal H_\text{env}$, where in turn $\mathcal H_ 
\text{sys} = \mathcal H_{\vect k} \otimes \mathcal H_{-\vect k}$.  
Notice that this separation does not correspond to the Fock space  
decomposition. The entire system is in a state $\hat \rho$; the state  
of the reduced system is $\hat\rho_\text{s} = \Tr_\text{env} {\hat\rho} 
$, and the state of the environment is $\hat\rho_\text{e} = \Tr_ 
\text{sys} {\hat\rho}$.  Generally speaking, the state for the entire  
system is not a factorised product state (\ie, $\hat \rho \neq \hat 
\rho_\text{s} \otimes \hat\rho_\text{e}$).

The action can be decomposed as $S = S_\text{sys} + S_\text{count} + S_ 
\text{env} + S_\text{int}$, where $S_\text{sys}$ is the renormalized  
system action,
\begin{subequations}\label{OQSAction}
\begin{equation}
	S_\text{sys} = \int \ud t  \left( \dot \phi_\vect p \dot \phi_{-\vect  
p} -E^2_\vect p \phi_\vect p \phi_{-\vect p} \right), \label{SysAction} 
\\
\end{equation}
$S_\text{count}$ is the appropriate counterterm action,
\begin{equation} \label{OQSActionEnv}
	S_\text{count}=
	\int \ud t  \left\{ ({\mathcal Z}_\vect p-1)\dot \phi_\vect p \dot  
\phi_{-\vect p} -\big[{\mathcal Z}_\vect p (\vect p^2+m_{0}^2)- E^2_ 
\vect p\big] \phi_\vect p \phi_{-\vect p} \right\},
\end{equation}
\end{subequations}
(with $m_0$ being the bare mass of the field), and $S_\text{env}$ and  
$S_\text{int}$ are the environment and interaction actions,  
respectively, which depend on the particular field theory model.
Notice that we have allowed for an arbitrary rescaling of the field $ 
\phi \to \phi/{\mathcal Z}^{1/2}_\vect p$ and for an arbitrary  
frequency of the two-mode system $E_\vect p$ [which needs not be  
necessarily of the form $(\vect p^2 + m^2)^{1/2}$].

Let us draw our attention on the field rescaling and the frequency  
renormalisation. Since it is always possible to freely move finite  
terms from the system to the counterterm action and \emph{vice versa},  
both the field rescaling and the frequency renormalisation should be  
taken into account even if no infinities appeared in the perturbative  
calculations. A physical criterion needs to be chosen in order to fix  
the values of these two parameters. In the vacuum, such criterion is  
provided by the on-shell renormalisation scheme; in non-vacuum  
situations, it is investigated in Ref.~\cite{Arteaga08a}, and will be briefly discussed in Sect.~\ref{sect:application}. Notice that   
the form of ${\mathcal Z}_\vect p$ and  $E_\vect p$ is not necessarily  
dictated by the Lorentz symmetry: even if the countertems which remove  
the infinities from the vacuum theory  also remove the infinities in general  
field states, there can be finite Lorentz-breaking contributions.
Anyway for the purposes of most of this paper the values of ${\mathcal Z}_ 
\vect p$ and  $E_\vect p$ are not be relevant and will be left  
unspecified (some comments will be made in Sect.~\ref{sec:Master} though).

The system variables $\phi_\vect p$ and $\phi_{-\vect p}$ are complex  
quantities verifying $\phi_{-\vect p}^* = \phi_{\vect p}$. We can  
construct real degrees of freedom by introducing the following change  
of variables:
\begin{equation}\label{ChangeVar}
	\phi_{\Sigma} = \frac{1}{\sqrt{2}} (\phi_{\vect p} + \phi_{-\vect  
p}), \qquad
	\phi_\Delta = \frac{-i}{\sqrt{2}} (\phi_{\vect p} - \phi_{-\vect p}).
\end{equation}
In terms of the real variables, the system action can be reexpressed as
\begin{equation}
	S_\text{sys} = \frac 12\int \ud t \big( \dot \phi_{\Sigma}^2 - E_ 
\vect p^2 \phi_{\Sigma}^2 + \dot \phi_\Delta^2 - E_\vect p^2 \phi_ 
\Delta \big).
\end{equation}
We could alternatively have directly obtained these two real degrees  
of freedom by working with the sine and cosine  Fourier transform   
\cite{HuMatacz94}. However we prefer to work with the exponential  
Fourier transform to make manifest the momentum conservation properties.

Most information on the reduced quantum system can be extracted from  
the set of correlation functions, or equivalently from the CTP  
generating functional for the reduced system, which can be written as
\begin{equation} \label{GenFunct}
\begin{split}
	Z[j_{a,\alpha}]&=\exp\bigg[-\frac{1}{2!}\int \ud t \ud{t'} j_ 
\alpha^{a}(t) j_\beta^{b}(t) G^{\alpha\beta}_{ab}(t,t')\\ &\qquad- 
\frac{1}{4!}\int \ud t \ud{t'}\ud{t''} \ud{t'''} j_\alpha^{a}(t) j_ 
\beta^{b}(t') j_\gamma^c(t'') j_\delta^d(t''') G^{\mathrm{(C)}\alpha 
\beta\gamma\delta}_{abcd}(t,t',t'',t''') + \cdots\bigg].
\end{split}
\end{equation}
This expression is somewhat cumbersome and needs some clarification.  
Latin indices $a,b,c\ldots$ are CTP indices and take the values 1 and  
2, indicating respectively the forward and backwards time branches    
characteristic of the CTP formalism (see appendix \ref{app:CTP}).  
Greek indices $\alpha,\beta,\gamma,\ldots$ take the two values $+\vect  
p$ and $-\vect p$, and make reference to the two field modes $\phi_ 
\vect p$ and $\phi_{-\vect p}$. An Einstein summation convention is  
used both for Latin and Greek indices. The propagator $G^{\alpha 
\beta}_{ab}(t,t')$ is the 2-point propagator connecting CTP indices $a 
$ and $b$, whose external legs correspond to particles with momenta $ 
\alpha$ and $\beta$. When the state is translation-invariant and  
isotropic, momentum conservation imposes:
\begin{equation}
\begin{split}
	G^{(+\vect p)(-\vect p)}_{ab}(t,t') &= G^{(-\vect p)(+\vect p)}_{ab} 
(t,t') =  G_{ab}(t,t';\vect p), \\
	G^{(+\vect p)(+\vect p)}_{ab}(t,t') &= G^{(-\vect p)(-\vect p)}_{ab} 
(t,t') =  0.
\end{split}
\end{equation}
In turn, $G^{\mathrm{(C)}\alpha\beta\gamma\delta}_{abcd}(t,t',t'',t''') 
$ is the connected part of the four-point correlation function having   
external legs with momenta $\alpha$, $\beta$, $\gamma$ and $\delta$.  
For translation-invariant states momentum conservation implies that  
only when momentum is balanced (\ie, two incoming and two outgoing  
external legs) the correlation function is non-vanishing. Terms with a  
higher number of external legs behave similarly.

The open quantum system is non-linear, and a systematic treatment of  
the generating functional can be done by using the tools of non-linear  
open quantum systems. We follow a different path in the next section.

\section{Quantum Brownian motion analogy}\label{sect:QBManalogy}

Let us reconsider the generating functional, \Eqref{GenFunct}. It  
depends on the $n$-point correlation functions, with $n$ being  
arbitrarily large. However, in many situations one is interested in  
properties which only depend on the two-point correlation functions.  
In other cases one is doing a perturbative expansion of the generating  
functional, and connected higher order correlation functions are  
usually also of higher order in the expansion parameter.  Finally  
there are situations in which one only has access to the two-point  
correlation functions, and expects (or simply hopes) that connected  
higher order correlation functions are subdominant. In any of this  
situations one can be tempted to approximate the generating functional  
by the following Gaussian expression:
\begin{equation}
\begin{split}
	Z[j_{a,\alpha}]&\approx\exp\bigg[-\frac{1}{2!}\int \ud t \ud{t'} j_ 
\alpha^{a}(t) j_\beta^{b}(t) G^{\alpha\beta}_{ab}(t,t') \bigg].
\end{split}
\end{equation}
This Gaussian approximation can be controlled in the framework or the  
large-$N$ expansion (where $N$ is the number of scalar fields)  
\cite{Tomboulis77,CopperMottola87,CooperEtAl94}. We should emphasise  
that it does not necessarily imply any perturbative expansion in the  
coupling parameter, nor any free field approximation.

The Gaussian approximation of the generating functional can be  
expanded to
\begin{equation} \label{ZCTPGaussian}
\begin{split}
	Z[j_{a,\vect p},j_{a,-\vect p}]&=\exp\bigg\{-\frac 12\int \ud t  
\ud{t'}\left[ j_{\vect p}^{a}(t) j_{-\vect p}^{b}(t') G_{ab}(t,t') +   
j_{\vect p}^{a}(t') j_{-\vect p}^{b}(t) G_{ab}(t,t')\right] \bigg\},
\end{split}
\end{equation}
or equivalently, in terms of the real variables $\phi_{\Sigma}$ and $ 
\phi_{\Delta}$ and its corresponding classical sources,
\begin{equation}
\begin{split}
	Z[j_{a,\Sigma},j_{a,\Delta}]&=\exp\bigg\{-\frac 12 \int \ud t \ud{t'}  
\left[ j_{\Sigma}^{a}(t) j_{\Sigma}^{b}(t') G_{ab}(t,t') +  
j_{\Delta}^{a}(t) j_{\Delta}^{b}(t') G_{ab}(t,t') \right] \bigg\}.
\end{split}
\end{equation}
or
\begin{equation} \label{ZDirectBasis}
	Z[j_{a,\Sigma},j_{a,\Delta}] = Z[j_{a,\Sigma}] Z[j_{a,\Delta}],  
\qquad  Z[j_{a}] = \exp{\bigg[-\frac 12 \int \ud t \ud{t'}  j^{a}(t)  
j^{b}(t') G_{ab}(t,t')  \bigg]}.
\end{equation}
Eq.~\eqref{ZDirectBasis} shows that for translation-invariant states  
within the Gaussian approximation the reduced two-mode state  
effectively behaves as two decoupled quantum mechanical degrees of  
freedom.

The Gaussian approximation implies that  the expression of the  
generating functional of a reduced two-mode in terms of the two-point  
propagators coincides with that of a QBM model, provided the system is  
isotropic and translation-invariant: compare \Eqref{ZQBMDirectBasis}  
with \Eqref{ZDirectBasis}. Moreover, by comparing Eqs.~\eqref{GOtherFourier} with Eqs.~\eqref{PropsSigma}, we also realise  
that the structure of the two-point propagators is identical in both  
cases. Notice that this latter fact is independent of the Gaussian  
approximation.

Therefore, we conclude that, assuming homogeneity, isotropy and  
stationarity, \emph{there is an equivalent QBM system for every scalar two-mode pair treated under the Gaussian approximation}. In other words,
within the Gaussian approximation, every two-mode of a given quantum  
field theory can be described in terms of  a pair of quantum Brownian  
particles interacting linearly with some effective environment. We  
must stress that, similarly as the linear interaction does not  
coincide with  the real coupling, the effective environment  does not  
coincide  with the real environment. The precise details of the  
equivalence are summarised in table\ \ref{tbl:equivalence} for the  
particular case of the $\lambda\phi^4$ theory. We will see below that  
both the strength of the linear coupling and the state of the  
equivalent environment depend on the details of the original  
environment.

\begin{table}
\centering
\begin{tabular}{lcc}
\hline\hline
& Original system 				& Equivalent linear QBM\\ \hline
System & two field modes   & two identical oscillators \\
System d.o.f. & $\phi_\vect p$, $\phi_{-\vect p}$ & 	2 copies of $q$ \\

Environment & other modes 3-d field  & 2 1-d field\\
Env. d.o.f. & $\phi_\vect q$, $\vect q \neq \pm\vect p$ & $\varphi_p$   
	\\

Frequency & $E_\vect p$ & $\Omega$ ($=$) \\

Coupling & $\frac{\lambda}{3!} \sum_{\vect q\vect q'}
		\phi_{\pm\vect p} \phi_{\vect q} \phi_{\vect q'} \phi_{\mp\vect p- 
\vect q -\vect q'}$  & $\glin^2 \sum_p \mathcal I(p) \dot q \varphi_p$\\
2-point function & $G_{ab}(t,t';\vect p)$ & $G_{ab}(t,t')$ $(=)$\\
\hline\hline
\end{tabular}
\caption{Detail of the equivalent linear QBM system for a $\lambda  
\phi^4$ quantum field theory. The $\lambda \phi^4$ model has been  
chosen for concreteness, but the correspondence would be analogous for  
any other field theory model. The symbol $(=)$ indicates that the  
original and equivalent quantities are indeed identical despite the  
name change. } \label{tbl:equivalence}
\end{table}


Let us investigate on this correspondence.
On the one hand, as is is shown in appendix \ref{app:QBM}, the effect  
of the environment in the QBM system is fully encoded in two kernels,  
the dissipation kernel $D(t,t')$ ---or its closely related counterpart  
$H(t,t')$, see \Eqref{kernelH}--- and the noise kernel $N(t,t')$.  The  
dynamics of the quantum Brownian particle can be determined once the  
frequency  and the noise and dissipation kernels are known. On the  
other hand, as shown in appendix \ref{app:CTP}, in a quantum field  
theory the two-point correlation functions are fully characterised by  
the frequency $E_\vect p$ and the retarded and Hadamard self-energies,  
$\SigmaR(t,t')$ and $\Sigma^{(1)}(t,t')$ respectively. By comparing  
again Eqs.~\eqref{GretFourier} and  \eqref{GOtherFourier} with Eq.\  
\eqref{PropsSigma} we realise that the precise analogy goes as follows:
\begin{subequations} \label{equivalenceQBM}
\begin{align}
	 \Omega &= E_\vect p ,\\
	H(t,t')&=\vect p^2 + m^2 - E_\vect p + \SigmaR(t,t';\vect p)  ,\\
	D(t,t') &= -i\Im \SigmaR(t,t';\vect p),\\
	N(t,t') &= \frac{i}{2} \SigmaN(t,t';\vect p) ,
\end{align}
\end{subequations}
where the quantities on the right hand side correspond to the original  
field theory system and the quantities on the left hand side  
correspond to the equivalent QBM model. The correspondence is valid no matter the  
renormalisation scheme chosen to fix $\mathcal Z_\vect p$ and $R_\vect  
p$.

This representation provides thus a first rough interpretation for the  
retarded and Hadamard self-energies, $\SigmaR(t,t';\vect p)$ and $ 
\SigmaN(t,t';\vect p)$ respectively. The retarded self-energy  
corresponds to the dissipation kernel, so that it determines the  
dissipative properties of the system, and it is independent of the  
state of the equivalent environment (though not independent of the  
state of the original environment).  The Hadamard self-energy  
corresponds to the noise kernel, and thus it is basically related to  
fluctuations.

\index{Dissipation kernel}
\index{Noise kernel}

Although the description in terms of the noise and dissipation kernels  
is often sufficient, the equivalent QBM system can be alternatively  
described in terms of the linear coupling constant to the effective  
environment, $\glin$, the distribution of frequencies of the  
environment $\mathcal I(\omega)$ (see appendix \ref{app:QBM}), and the  
occupation number of the modes of the effective environment $n(p) =  
\Tr [\hat\rho_\text{e}^\text{(eff)} \hat a_p^\dag \hat a_p ]$. The  
product $\glin^2 \mathcal I(\omega)$, which determines the coupling  
strength to the the $\omega$-mode of the effective environment, can be  
obtained from the imaginary part of the self-energy:
\begin{subequations} \label{CorrDetails}
\begin{equation}
	\Im \SigmaR(\omega;\vect p) = iD(\omega) = - \frac{  \glin^2}{2}  
\omega\mathcal I(\omega).
\end{equation}
This last equation implies that the equivalent coupling depends on the  
state of the real environment, since the retarded self-energy is state-dependent in general.
The occupation numbers of the effective environment $n(p)$ can be  
reproduced from the Hadamard self-energy:
\begin{equation}
	\Sigma^{(1)}(\omega;\vect p) = -2i N(\omega) = -2i\glin^2 |\omega|  
\mathcal I(\omega) \left[ \frac{1}{2} + n(|\omega|) \right],
\end{equation}
\end{subequations}
The knowledge of the occupation numbers fully determine a Gaussian  
stationary state for the equivalent environment. These results follow  
from Eqs.~\eqref{DisQBM} and \eqref{NoiseQBM}.

\index{Occupation number}

Note that the analogy depends only on the Gaussian approximation, so  
that it can be extended to all orders in perturbation theory. Notice  
also that the results derived from the QBM interpretation are exact  
for all those properties which depend only on the two-point  
correlation functions; for the properties which depend on higher order  
correlation functions, it is a correct approximation depending on the  
validity of the Gaussian approximation, \ie, depending on the relative  
importance of the connected parts of the correlation functions with  
respect to the disconnected parts.

\section{Example applications} \label{sect:application}

\subsection{Interpretation of the self-energy}

\index{Self-energy!imaginary part} 

As an first example application, let us make use of the QBM  
correspondence  to analyse the physical significance of the self-energy in general backgrounds. Our findings will coincide with the  
result by Weldon \cite{Weldon83} (which nowadays is a textbook result  
\cite{LeBellac}) as far as the imaginary part of the self-energy is  
concerned. However, while Weldon's original analysis was only valid to  
first order in perturbation theory, our technique is be valid to all  
orders. Moreover, our analysis will not be tied to any field theory  
model. Additionally we will also obtain an interpretation for the  
other components of the self-energy.  In this case the QBM analogy is  
exact since no four-point correlation functions are involved.

To start we consider the probability that an excitation of energy $ 
\omega$ decays into the one-dimensional environment, in the equivalent QBM system. The probability $\Gamma_-$ that an excitation of  
the Browian particle with positive energy $\omega$ decays into the  
environment is given by (see \eg\ Ref.~\cite{LeBellac})
\begin{equation}
	\Gamma_-(\omega) = \frac{1}{2\omega} \int \frac{\vd k}{2\pi2|k|}   
2\pi \delta(\omega-|k|) |\mathcal M|^2 [1+n(|k|)],
\end{equation}
where $\mathcal M$ is the amplitude of the transition and $n(|k|)$ is  
the occupation number of the environment states with energy $|k|$. The  
factor $1+n(|k|)$, which is due to the Bose-Einstein statistics,  
enhances the decay probability to those states which are  already  
occupied. Since the equivalent QBM system is linear, the squared decay  
amplitude is simply given to first order in the linear coupling  
constant $g$ by
\begin{equation} \label{matrixM}
	|\mathcal M|^2 = \glin^2 \mathcal I(\omega) \omega^2,
\end{equation}
where  $\mathcal I(\omega)$ is the distribution of frequencies of the  
effective environment. The factor $\omega^2$ is a consequence of the  
derivative coupling in the QBM model. The decay probability is  
therefore:
\begin{equation} \label{GammaMinus}
	\Gamma_-(\omega) = \frac{1}{2} \glin^2 \mathcal I(\omega)  
[1+n(\omega)].
\end{equation}
Likewise, the probability that an excitation of positive energy $\omega 
$ is created spontaneously from the environment is given by
\begin{equation} \label{GammaPlus}
	\Gamma_+(\omega) = \frac{1}{2\omega} \int \frac{\vd k}{2\pi2|k|}   
2\pi \delta(\omega-|k|) |\mathcal M|^2 n(|k|) = \frac{1}{2} \glin^2  
\mathcal I(\omega) n(\omega).
\end{equation}
In the original system, $\Gamma_-$ can be interpreted as the  
probability that a (possibly off-shell)  excitation with energy $\omega 
$ decays into the environment, and $\Gamma_+$ can be interpreted as  
the probability that an environment spontaneously creates an  
excitation with energy $\omega$.

Notice that the notion of decay rate in quantum mechanics is  
meaningful only when the excitations are long-lived, or, to put it  
differently, when the product $\glin^2 \mathcal I(\omega) $ is very  
small. There is an inherent uncertainty in the concept of decay rate,   
which can be traced to the time-energy uncertainty principle.  
Therefore it is sufficient to present results to first order in $g$:  
the inherent uncertainty to the notion of decay rate in quantum  
mechanics  is of the same order as the error done by neglecting   
higher powers of $g$. In any case, this does not mean that we are  
doing any perturbative expansion in the original system: the decay  
rate can be computed to any desired order in the original perturbative  
coupling constant (\eg, $\lambda$ in the case of the $\lambda\phi^4$  
theory).

\index{Decay rate}
We next analyse the self-energy components in the equivalent QBM mode.  
Given that $\Im \SigmaR(\omega) = (i/2)[\Sigma^{21}(\omega) -  
\Sigma^{12}(\omega)]$ (see appendix \ref{app:CTP}), we start by  
analyzing $\Sigma^{21}(\omega)$. Applying CTP Feynman rules we get
\begin{equation}
	-i\Sigma^{21} (t,t')= - (i\glin)^2 \partial_t\partial_{t'}  \int  
\frac{\vd p}{2\pi} \mathcal I(p) \Tr_\text{env}{\big[\hat\rho_\text{e}  
\hat \varphi_{\text Ip}(t) \hat \varphi_{\text I(-p)}(t')\big]},
\end{equation}
where $\hat \varphi_{\text Ip}$ is $p$-mode of the environment field   
in the interaction picture.  We have exploited that first  
order perturbation theory yields exact results for the self-energy in  
linear systems. Introducing two resolutions of the identity in the  
basis of eigenstates of the environment Hamiltonian it is a simple  
exercise \cite{ArteagaThesis} to show that  the above equation can be  
developed to
\begin{equation*}
\begin{split}
	\Sigma^{21} (t,t')
&= i\glin^2  \partial_t \partial_{t'} \int_{0}^\infty \vd{p}  
\frac{\mathcal I(p)}{p}  \sum_n \rho_{p,n} \left[ (n+1) \expp{-i p(t- 
t')} +  n \expp{ip(t-t')} \right],
\end{split}
\end{equation*}
where $\rho_{p,n}$ is the diagonal value of the reduced density matrix  
of the $p$-mode of the equivalent environment: $ \Tr_{q\neq p} \langle  
n_p | \hat \rho_\text{e} |m_p\rangle = \rho_{p,n} \delta_{nm}$.  
Introducing the Fourier transform we get
\begin{equation*}
\begin{split}
	\Sigma^{21} (\omega)
&=  i\glin^2  \omega^2 \int_{0}^\infty \ud p \frac{\mathcal I(p)}{p}   
\sum_n \rho_{p,n} \left[ (n+1) \delta(\omega-p) +  n \delta(\omega+p)  
\right],
\end{split}
\end{equation*}
and restricting to positive energies,
\begin{equation}
	\Sigma^{21} (\omega)
= - i\glin^2  \omega  \mathcal I(\omega) \sum_n \rho_{\omega,n} (n+1)  
=  i\glin^2  \omega  \mathcal I(\omega) \left[1+n(\omega)\right],  
\quad \omega > 0,
\end{equation}
where we recall that $n(\omega) = \Tr [\hat\rho_\text{e}^\text{(eff)}  
\hat a_\omega^\dag \hat a_\omega ]  = \sum_n \rho_{\omega,n} n$ is the  
occupation number the $\omega$-mode of the environment. By comparing  
with \Eqref{GammaMinus}, we thus see that i proportional to the decay  
rate:
\begin{equation}
	\Sigma^{21}(\omega) = 2i\omega \Gamma_-(\omega), \quad \omega > 0.
\end{equation}

Repeating the calculation for  $\Sigma^{21}(\omega)$ we similarly find
\begin{equation}
	\Sigma^{12} (\omega)
=  i\glin^2  \omega  \mathcal I(\omega) \sum_n \rho_{\omega,n} n = i  
\glin^2 \omega \mathcal I(\omega) n(\omega), \quad \omega > 0,
\end{equation}
having the corresponding interpretation in terms of the creation rate,
\begin{equation}
	\Sigma^{12}(\omega) = 2i\omega \Gamma_+(\omega), \quad \omega > 0.
\end{equation}
When the energies are negative $\Sigma^{21}(\omega)$ and $\Sigma^{12} 
(\omega)$ exchange roles.

\index{Self-energy!retarded}
The imaginary part of the retarded self-energy is therefore given by
\begin{equation}
	\Im\SigmaR (\omega) = \frac{i}{2} [\Sigma^{21}(\omega) -\Sigma^{12} 
(\omega)]  = -\frac{1}{2} \glin^2  \omega  \mathcal I(\omega) ,
\end{equation}
and can be interpreted as the net decay rate for an excitation of  
energy $\omega$ ---\ie, decay rate minus creation rate:
\begin{equation} \label{Weldon}
\begin{split}
	\Im\SigmaR(\omega) = -\omega[\Gamma_-(\omega) - \Gamma_+(\omega)].
\end{split}
\end{equation}
We therefore recover Weldon's result \cite{Weldon83}.

\index{Self-energy!Hadamard}
We can additionally get an interpretation for the Hadamard self-energy. It is given by
\begin{equation} \label{SigmaNInterp1}
	\SigmaN(\omega)= - \Sigma^{21}(\omega) - \Sigma^{12}(\omega) = - i   
\glin^2  |\omega|  \mathcal I(\omega) [1+2n(|\omega|)],
\end{equation}
and is proportional to the probability of decay plus the probability  
of creation,
\begin{equation} \label{SigmaNInterp}
	\SigmaN(\omega)=-  2i|\omega|[\Gamma_-(\omega) + \Gamma_+(\omega)].
\end{equation}

\subsection{Master equation for relativistic quasiparticles}\label{sec:Master}

As another example application of the correspondence, let us study the master equation for the second-quantized relativistic quasiparticles. A quasiparticle an elementary excitation propagating in some  
backgroung, characterized by  a momentum $\vect p$, an energy  
$E_{\vect p}$, and a decay rate $\gamma_\vect p$ verifying $\gamma_\vect p  
\ll E_\vect p$ (so that the excitation is long-lived). The basic properties characterizing the long-time evolution of the relativistic  
quasiparticles can be extracted from the analysis of the retarded  
propagator corresponding to the field mode in a model-independent way \cite{Arteaga08a} (differently to the short-time behaviour, which is model-dependent). In particular, the real  
and imaginary parts of the self-energy, when evaluated on shell,  
correspond to the physical energy and the decay rate of the quasiparticle,  
respectively:
\begin{equation}
	E^2_{\vect p} = m^2 + \vect p^2 + \Re \SigmaR(E_{\vect p},\vect p),  
\qquad
	\gamma_{\vect p} = -\frac{1}{E_{\vect p}} \Im \SigmaR(E_{\vect p}.
\vect p),
\end{equation}
Using the QBM correspondence, these two equations can be equivalently written as:
\begin{equation}
	  \Re H(\Omega) = 0 , \qquad
	\gamma_{\vect p} = -\frac{i}{\Omega} D(\Omega),
\end{equation}
The value of $E_\vect p$ (or equivalently $\Omega$) is fixed by requiring that $E_\vect p$ represents the physical energy of the quantum mode as would be measured by a particle detector. We refer to Ref.~\cite{Arteaga08a} (which makes use of the QBM correspondence) for further details on the field theory analysis of the quasiparticle excitations.

Thus, let us study the master equation \eqref{master} for $q=\{\phi_\Sigma,\phi_\Delta\}$. To this end it will prove useful to express the on-shell values of the noise and dissipation kernels in terms of the energy $\Omega=E_\vect p$ and the decay rate $\gamma = \gamma_\vect p$:
\begin{equation}
	D(\Omega) = - H(\Omega) = i\gamma, \qquad N(\Omega) = \gamma \Omega  \left(\frac12 + n\right),
\end{equation}
where $n=n(|\Omega|)$ is the equilibrium occupation number of the mode in question. Notice that since the decay rate is necessarily small, the coupling constant to the effective environment is also small. Therefore the mode can be considered to be weakly coupled to the effective environment.

In the weak coupling regime, the master equation coefficients are given by Eqs.~\eqref{mastercoeff}, and are usually divergent for short times \cite{HuPazZhang92}. These divergences are associated to the fact that the assumption of factorized initial conditions for the inital state is unphysical, as commented in appendix \ref{app:QBM}. Since we are interested in the long-time behaviour of the quasiparticles, we study the asymptotic value of these coefficients, which is free from the short-time divergences.

It is easy to show that the frequency shift vanishes:
\begin{subequations} 
\begin{equation}
\begin{split}
	\delta \Omega^2  &=   -2 \int_{t_\text i}^\infty D(s,t_\text i) \cos{\Omega  
(s-t_{\text i})}\,  \vd s = \frac{1}{2} \Re \int_{-\infty}^\infty H(s,t_\text i) \expp{i\Omega  
(s-t_{\text i})}\,  \vd s = \Re H(\Omega) = 0.
\end{split}
\end{equation}
Taking into account that the dissipation kernel is antisymmetric, the evaluation of the dissipative factor is also straightforward:
\begin{equation}
\begin{split}
	\Gamma &=  \frac{1}{\Omega}  \int_{t_\text i}^\infty D(s,t)  
\sin{\Omega (s-t_{\text i})}\, \vd s =  \frac{1}{2\Omega i}   \int_{-\infty}^\infty D(s,t)  
\expp{i\Omega (s-t_{\text i})}\, \vd s \\ &= \frac{1}{2\Omega i}  D(\omega) = \frac{\gamma_\vect p}{2} .
\end{split}
\end{equation}
The first diffusion factor is also easily computed, this time recalling that the noise kernel is symmetric:
\begin{equation}
\begin{split}
	\Gamma h &= \int_{t_\text i}^\infty N(s,t_\text i) \cos{\Omega (s- 
t_{\text i})} \, \vd s = \frac12 \int_{-\infty}^\infty N(s,t_\text i) \expp{i\Omega (s- 
t_{\text i})} \, \vd s \\ &=\frac{N(\Omega)}{2} = { \Omega \gamma} \left(\frac12+n\right).
\end{split}
\end{equation}
The second diffusion factor requires some extra work:
\begin{equation*}
\begin{split}
	\Gamma f &= \frac{1}{\Omega} \int_{t_\text i}^\infty N(s,t_\text i)  
\sin{\Omega (s-t_{\text i})} \,\vd s=
	\frac{1}{\Omega} \Im \int_{-\infty}^\infty N(s,t_\text i) \theta(s-t_\text i)  
\expp{i\Omega (s-t_{\text i})} \,\vd s
	\\ &= \frac{1}{\Omega} \Im \int \frac{\vd \omega}{2\pi} \frac{ iN(\omega)}{\Omega-\omega+i\epsilon}
	=- \frac{1}{\Omega}  \text{P} \int \frac{\vd \omega}{2\pi} \frac{ iD(\omega) \sign(\omega)}{\Omega-\omega} [1+2n(|\omega|)].
\end{split}
\end{equation*}
The integrand in the last equality is only signifficantly different from zero when $\omega \sim \Omega$. Therefore as a first approximation we may write:
\begin{equation}
\begin{split}
	\Gamma f\approx -\frac{1}{\Omega}  \text{P} \int \frac{\vd \omega}{2\pi} \frac{-i D(\omega) }{\Omega-\omega} [1+2n(|\Omega|)] = \frac{1}{2\Omega}  \Re H(\Omega)  (1+2n)=0.
\end{split}
\end{equation}
\end{subequations}

Thus the master equation can be written as:
\begin{equation}
\begin{split}
	i \derp{}{t} \rho_\text s(q,q',t) &= \bigg[-\frac12 \left( \derp[2]{} 
{q} - \derp[2]{}{q'} \right) + \frac12 \Omega^2 (q^2-q'^2) \\
	&\quad  -  \frac{i}{2}\gamma(q-q') \left( \derp{}{q} - \derp{}{q'} \right) -  
i  \Omega \gamma\left(\frac12+n\right) (q-q')^2  \bigg] \rho_\text s(q,q',t).
\end{split}
\end{equation}
or, equivalently, in terms of the Wigner function, as\footnote{Notice that in this equation $p$ is the canonical momentum associated to the variable $q$, and has nothing to do with the physical quasiparticle momentum $\vect p$.}
\begin{equation}
	\derp{W_{\text s}}{t} = - p \derp{W_{\text s}}{q} + \Omega^2 \derp{W_{\text s}}{p} +\gamma
\derp{pW_{\text s}}{p}+ \frac{ \Omega \gamma}{4} \left(1+2n\right) \derp[2]{W_{\text s}}{p}.
\end{equation}
This is identical to the master equation found by Caldeira et al. \cite{CaldeiraEtAl89}, corresponding to weak couplings.\footnote{Fleming et al.~\cite{FlemingEtAl07} found logarithmically divergent results for the master equation coefficients in the ohmic dissipation case, even in the asymptotic regime and in the weak coupling limit. This divergences are associated to the unphysical ultraviolet behaviour of the ohmic dissipation model. Our approximate treatment overlooks the possible divergences associated to the high energy limit of the noise and dissipation kernel. }

The reduced density matrix and the Wigner function are to be interpreted in terms of the field modes, and not directly in terms of particles or quasiparticles themselves. Qualitatively, for Gaussian initial conditions the solution of the master equation can be described in terms of the cumulants of a Gaussian distribution in the following way  \cite{FlemingEtAl07,UnruhZurek89}. The expectation value of the field $\langle q \rangle$ follows the trajectory of a classical underdamped harmonic oscillator ($\ddot q + \gamma \dot q + \Omega^2 q = 0$), namely, it oscillates and slowly decays towards the origin at a rate $\gamma/2$. However, for single quasiparticle excitations the expectation value of the field is always vanishing. Therefore, it is more appealing to consider the dynamics of the second order cumulants, and in particular the dynamics of the energy of the mode, $(1/2) \langle \ddot q + \Omega^2 q \rangle$. When perturbed by the introduction of a quasiparticle, the energy of the mode slowly decays at a rate $\gamma$ towards its equilibrium value $\Omega(1/2 + n)$, with $n$ being the original occupation number of the mode \cite{Arteaga08a}.


\section{Summary and discussion}\label{sect:discussion}

In this paper  we have explored the open quantum system viewpoint for  
a pair of field modes of opposite momentum, which are the relevant  
degrees of freedom for the analysis of particle-like excitations in  
field theory. The main results have been, on the one hand, showing  
that in any interacting field theory, assuming homogeneity,  
stationarity and Gaussianity, this open quantum system can be  
equivalently represented by two identical quantum Brownian particles  
interacting {linearly} with an effective environment, and, on the  
other hand, exploring the details of the equivalence, which are  
expressed in Eqs.~\eqref{equivalenceQBM} and \eqref{CorrDetails} and  
in table \ref{tbl:equivalence}.

The Brownian motion equivalence is based on three simple  well-known  
observations. First, the fact that the structure of the two-point  
correlation functions, which we reexamined in the appendices, is  
identical regardless of the nature of the interactions. A byproduct  
of  this  fact has been establishing a link between the self-energy  
and the noise and dissipation kernels, thus connecting the open  
quantum system and field theory notation and languages. Second, the  
recognition that a Gaussian truncation leads to a generating  
functional which coincides with that of a linear theory. Finally, the  
observation that when assuming homogeneity, stationarity and isotropy  
the two-mode pair behave as a two copies of a single degree of freedom.

If the system is Gaussian, stationary and homogeneous, the equivalence  
is exact. If the system is non-Gaussian,  the analogy is also exact as  
far as two-point correlation functions are concerned, but only  
approximate for higher order correlation functions. If the system is  
non-stationary or non-homogeneous, there will be corrections to the  
results of the order of $Lp$, where $L$ is the characteristic  
inhomogeneity time or length scale, and $p$ is the relevant energy or  
momentum scale. Therefore the analogy is perfectly valid in non-homogeneous backgrounds as long as we consider modes whose 
characteristic wavelengths are much smaller than the inhomogeneity  
scale.

It must be noted that for the equivalence to be useful one still needs  
the field theoretic computation of the two-point correlation functions  
(or, equivalently, the self-energy). The correspondence does not help  
in this calculation, but must be regarded as a tool useful for  
interpreting and analyzing the dynamics of the two-mode system. In  
this sense, there are three basic characteristics which make the QBM  
equivalence appealing.

First, the equivalence is \emph{universal}, in the sense that it  
provides the most general description of the dynamics of the scalar  
two-mode pair within the Gaussian approximation. The description in  
terms of a linear open quantum system allows a unified description of  
many different quantum field theory systems: the details of the  
quantum field theory are unimportant once the equivalent noise and  
dissipation kernels are known. Different field theory models can be  
thus classified in the same equivalence class if they lead to the same  
QBM equivalent model.

Second, the equivalence provides a  \emph{simple} characterization of  
the interaction, given that an arbitrarily complicated coupling with  
any number of fields is reduced to a linear interaction with a one-
dimensional field. Linearly interacting systems have been thoroughly  
studied in the literature (see \eg\ references given in the  
introduction and appendix \ref{app:QBM}), and exhibit many intersting  
properties, among which one can cite the fact that they are exactly  
solvable. Therefore, within the regime of validity of the Gaussian  
approximation, all the methods developed in the literature for linear  
quantum systems can be applied to analyse the dynamics of a two-mode  
pair.

Finally, the equivalence is \emph{non-perturbative}, since it does not  
rely on any perturbative development in the coupling constant of the  
original field theory. It can be therefore extended to all orders in  
perturbation theory and applied to strongly interacting systems.

The QBM correspondence, which has also been implicitly used in Refs.~\cite{Parentani07a,Parentani07b}, proves a useful tool when analyzing  
the properties of particle-like excitations in general backgrounds  
from a field theory perspective \cite{Arteaga08a,Arteaga08b}.
The utility of the QBM analogy is already highlighted by the first two immediate  
applications presented in this paper. 

As a first application, we reexplored the well-known  
result that the imaginary part of the retarded self-energy   
corresponds to the net decay rate of the particle excitations. This  
was a rather simple application which however improved the  
conventional textbook derivation by taking profit of the three   
properties we have remarked above: in the first place, the universality of the QBM  
analogy, given that the derivation presented in this paper was not  
tied to any specific field theory model; in the second place, its simplicity, since  
the calculation essentially reduced to doing trivial perturbative  
expansions in the equivalent linear system, and, finally, its non-perturbative character, since the derivation avoided any perturbative  
expansion in the original system.

As a second application, we presented the relevant master equation for the dynamics of the modes corresponding to the quasiparticle momentum.  This was a slighly more involved application, relying on the results of Ref.~\cite{Arteaga08a} (and also complementing them), which highligted the fact that the linear open quantum system machinery can  readly  be exported to quantum field theory.
The presentation and analysis of the master equation, as done in the paper must be understood as a first approximation to the problem, which might be sufficiently interesting by itself to deserve further work.


\acknowledgments{I am very grateful to Albert Roura for several helpful discussions  
and suggestions, and to Renaud Parentani and Enric Verdaguer for their  
comments on the original manuscript. This work is partially supported  
by the Research Projects MEC
	FPA2007-66665C02-02 and DURSI 2005SGR-00082.}

\appendix
\section{The closed time path method and the two-point propagators in  
field theory} \label{app:CTP}

\index{Closed time path method|(}
\index{In-in method|see{closed time path method}}
\index{CTP|see{closed time path method}}

In this appendix we give a brief introduction to the closed time path  
(CTP) method (also called \emph{in-in} method, in contrast to the  
conventional \emph{in-out} method), stressing those aspects relevant  
for this paper, and apply it to the analysis of the two-point  
propagators. We address the reader to Refs.~\cite{ChouEtAl85,CalzettaHu87,CamposHu98,CamposVerdaguer94,Weinberg05}  
for further details on the CTP method, and to  Refs.~\cite 
{DasThermal,HardmanEtAl87,Lawrie88,LawrieMcKernan96,ArteagaThesis} for  
further details on the structure of the two point functions.
For the purposes of this appendix we shall consider a free or an  
interacting scalar field $\phi$, although results also apply for a  
single quantum mechanical degree of freedom.

The path-ordered generating functional $Z_{\mathcal C}[j]$ is defined
as
\begin{equation}
   Z_{\mathcal C}[j] = \Tr \left(\hat \rho T_{\mathcal C} \expp{i  
\int_{\mathcal C} \vd t \int
   \ud[3]{\vect
   x} \hat \phi(x) j(x)}  \right),
\end{equation}
where $\hat \phi(x)$ is the field operator in the Heisenberg
picture, ${\mathcal C}$ is a certain path in the complex $t$
plane, $T_{\mathcal C}$ means time ordering along this path and $j(x)$
is a classical external source.  By functional differentiation of
the generating functional with respect to $\phi$, path-ordered
correlation functions can be obtained:
\begin{equation}
	G_\mathcal C(x,x') = \Tr\big[ \hat\rho T_{\mathcal C} \hat\phi(x) \hat 
\phi(x')\big] =  - \left. \frac{ \delta^2  Z_{\mathcal C}}{\delta j(x)  
\delta j(x')}\right|_{j=0}.
\end{equation}
Introducing a complete basis of eigenstates of the field operator in  
the Heisenberg picture,
$	\hat\phi(t,\vect x) |\phi,t\rangle = \phi(t,\vect x) |\phi,t\rangle 
$, as a representation of the identity, the generating functional can  
be expressed as:
\begin{equation}  \label{PrePathGenFunct}
	Z_\mathcal C[j] = \int \widetilde{\mathrm d} \phi\,  
\widetilde{\mathrm d} \phi' \langle \phi,\ti|\hat\rho|\phi',\ti\rangle
	\langle \phi',\ti |T_{\mathcal C} \expp{i \int_{\mathcal C} \vd t \int
   \ud[3]{\vect
   x} \hat \phi(x) j(x)}  |\phi,\ti\rangle
\end{equation}
The functional measures $\widetilde{\mathrm d} \phi$ and $ 
\widetilde{\mathrm d} \phi'$ go over all field configurations of the  
fields at fixed initial time $t$. If the path $\mathcal C$ begins and  
ends at the same point $\ti$, then the transition element of the  
evolution operator can be computed via a path integral:
\begin{equation}\label{PathGenFunct}
	Z_\mathcal C[j] = \int \widetilde{\mathrm d} \phi\,  
\widetilde{\mathrm d} \phi' \langle \phi,\ti|\hat\rho|\phi',\ti\rangle
	\int_{\varphi(\ti,\vect x)=\phi(\vect x)}^{\varphi(\ti,\vect x)= 
\phi'(\vect x)} \mathcal D \varphi \expp{i \int_{\mathcal C} \vd t \int
   \ud[3]{\vect x}\{ L[\varphi] +  \varphi(x) j(x)\}},
\end{equation}
where $L[\phi]$ is the Lagrangian density of the scalar field.

\begin{figure}
   \centering
   \includegraphics{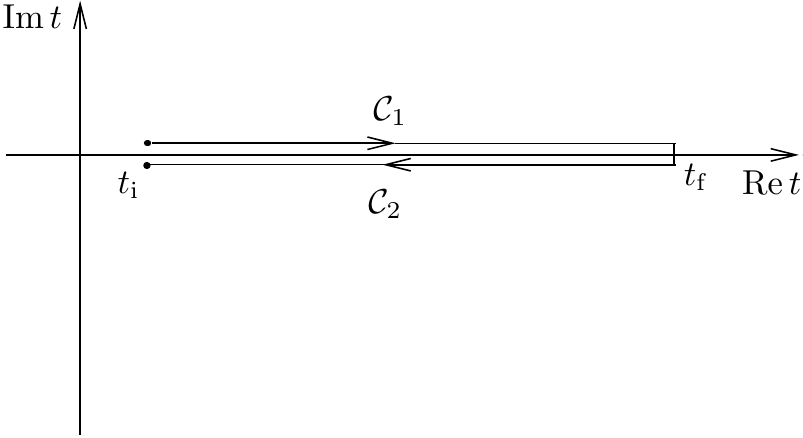}
   \caption{Integration path in the complex-time plane used in the
   CTP method. The forward and backward lines are infinitesimally  
close to the real axis.}
   \label{fig:CTP}
\end{figure}

Let us consider the time path shown in Fig.~\ref{fig:CTP}.  If we  
define $\varphi_{1,2}(t,\vect
x)=\varphi(t,\vect x)$ and $j_{1,2}(t,\vect x)=j(t,\vect x)$ for $t
\in {\mathcal C}_{1,2}$, then the generating functional can be  
reexpressed as:
\begin{equation}	 \label{CTPGenFunct}
\begin{split}
	Z[j_1,j_2] &= \int \widetilde{\mathrm d} \phi\, \widetilde{\mathrm d}  
\phi' \widetilde{\mathrm d} \phi'' \langle \phi,\ti|\hat\rho|\phi',\ti 
\rangle  \\
	&\qquad\times\int_{\varphi_1(\ti,\vect x)=\phi(\vect  
x)}^{\varphi_1(\tf,\vect x)=\phi''(\vect x)} \mathcal D \varphi_1
	\expp{i \int \ud[4]{x}\{ L[\varphi_1] +  \varphi_1(x) j_1(x)\}}\\
	&\qquad\times\int_{\varphi_2(\ti,\vect x)=\phi'(\vect  
x)}^{\varphi_2(\tf,\vect x)=\phi''(\vect x)} \mathcal D \varphi_2
	\expp{-i\int \ud[4]{x}\{ L[\varphi_2] +  \varphi_2(x) j_2(x)\}}.
\end{split}
\end{equation}
In the following it will prove useful to use a condensed notation  
where neither the boundary conditions of the path integral nor the  
integrals over the initial and final states are explicit. With this  
simplified notation the above equation becomes
\begin{equation}	
\begin{split}
	Z[j_1,j_2] &=  \int \mathcal D \phi_1 \mathcal D \phi_2\langle \phi,t| 
\hat\rho|\phi',t\rangle
	\expp{i \int \ud[4]{x}\{ L[\varphi_1] - L[\varphi_2] +  \varphi_1(x)  
j_1(x)-  \varphi_2(x) j_2(x)\}}
\end{split}
\end{equation}
An operator representation is also possible:
\begin{equation}\label{ZCTPOper}
   Z[j_1,j_2] = \Tr
	\left(\hat \rho \,
	\widetilde T \expp{-i \int_\ti^\tf \vd t \int
   \ud[3]{\vect
   x} \hat \phi(x) j_2(x)} T \expp{i \int_\ti^\tf \vd t \int \ud[3] 
{\vect
   x} \hat \phi(x) j_1(x)} \right).
\end{equation}

\index{Propagator!Feynman}
\index{Propagator!Dyson}
\index{Propagator!Whightman}

By functionally differentiating the generating functional, the Feynman  
and Dyson propagators and the Whightman functions can be obtained:
\begin{subequations} \label{CTPPropSet}
\begin{align}
	G_{11}(x,x') &= \GF (x,x') = \Tr{\big[ \hat\rho T \hat\phi(x) \hat 
\phi(x')\big]} =  - \left. \frac{ \delta^2  Z}{\delta j_1(x) \delta  
j_1(x')}\right|_{j_1=j_2=0}, \\
	G_{21}(x,x') &= G_+ (x,x') = \Tr\big[ \hat\rho \hat\phi(x) \hat 
\phi(x')\big] =   \left. \frac{ \delta^2  Z}{\delta j_2(x) \delta  
j_1(x')}\right|_{j_1=j_2=0} ,\\
	G_{12}(x,x') &= G_- (x,x') = \Tr\big[ \hat\rho \hat\phi(x') \hat 
\phi(x)\big] =  \left. \frac{ \delta^2  Z}{\delta j_1(x) \delta  
j_2(x')}\right|_{j_1=j_2=0} ,\\
	G_{22}(x,x') &= G_\text{D} (x,x') = \Tr\big[ \hat\rho \widetilde T  
\hat\phi(x) \hat\phi(x')\big] =  - \left. \frac{ \delta^2  Z}{\delta  
j_2(x) \delta j_2(x')}\right|_{j_1=j_2=0}.
\end{align}
\end{subequations}
These four propagators can be conveniently organised in a $2\times2$  
matrix, the so-called direct matrix.
\begin{equation}  \label{DirectBasisCTP}
	G_{ab}(x,x') = \begin{pmatrix}
				\GF (x,x') & G_-(x,x')\\
				G_+ (x,x') & G_\text{D} (x,x')
	              \end{pmatrix}
\end{equation}
Lowercase roman indices may acquire the values 1 and 2 are raised and  
lowered with the ``CTP metric''
$c_{ab}=\mathrm{diag}(1,-1)$. Higher order correlation functions can  
be obtained in a similar way.

We may also consider the Pauli-Jordan or commutator propagator,
\index{Propagator!Pauli-Jordan}
\begin{subequations}
\begin{equation}
G(x,x') = \Tr{\big(
\hat\rho\,[  \hat\phi(x) ,\hat\phi(x')] \big)
},
\end{equation}
and the Hadamard or anticonmutator function
\index{Propagator!Hadamard}
\begin{equation}
	G^{(1)}(x,x') = \Tr{\big(
\hat\rho\,\{  \hat\phi(x) ,\hat\phi(x') \} \big)}.
\end{equation}
\end{subequations}
For linear systems\footnote{By linear systems we mean systems whose  
Heisemberg equations of motion are linear. These correspond either to  
non-interacting systems or to linearly coupled systems.} the Pauli-Jordan propagator is independent of the state and carries information  
about the system dynamics. Finally, one can also consider the retarded  
and advanced propagators
\index{Propagator!retarded}
\index{Propagator!advanced}
\begin{subequations}
\begin{align}
   G_\mathrm {R}(x,x') &= \theta(x^0-x'^0) G(x,x') =\theta(x^0-x'^0)  
\Tr{\big(
\hat\rho\, [\hat \phi(x),\hat\phi(x')]\big)}, \label{RetardedProp}\\
   G_\mathrm {A}(x,x') &= \theta(x'^0-x^0) G(x,x') =\theta(x'^0-x^0)  
\Tr{\big(
\hat\rho\, [\hat \phi(x),\hat\phi(x')]\big)},
\end{align}
\end{subequations}
which also do not depend on the the state for linear systems. The  
retarded and advanced propagators and the Hadamard function can be  
used as an alternative basis to \eqref{DirectBasisCTP} in the so-called physical or Keldysh basis.

It is also useful to introduce the correlation functions in momentum  
space, which are defined as the Fourier transform of the spacetime  
correlators with respect to the difference variable $\Delta=x-x'$  
keeping constant the central point $X=(x+x')/2$:
\begin{equation}
   G_{ab}(\omega,\vect p;X) = \int \ud[4]{\Delta} \expp{i \omega  
\Delta^0 - i\vect p\cdot \vect \Delta}
   G_{ab}(X+\Delta/2,X-\Delta/2).
\end{equation}
Mixed time-momentum representations of the propagator, $G_{ab}(t,t'; 
\vect p;\vect X)$,  can be similarly introduced.
For homogeneous and static backgrounds the Fourier transformed  
propagator does not depend on the mid point $X$. The canonical example  
of static and homogeneous background is the thermal background, in  
which the state of the field is
$	\hat\rho = {\expp{- \beta \hat H}}/{\Tr{\big(\expp{- \beta \hat H} 
\big)}}$.
Thermal field theory  can be thus treated as a particular example of  
field theory over an arbitrary background. This viewpoint corresponds  
to the so-called real time approach to field theory  
\cite{DasThermal,LeBellac,LandsmanWeert87}.

For interacting theories the self-energy can be introduced similarly  
to the vacuum case. Interaction theory mixes the two CTP branches and  
therefore the self-energy has matrix structure and is implicitly  
defined through the equation
\begin{equation} \label{SelfEnergyGeneral}
   G_{ab}(x,x') = G_{ab}^{(0)}(x,x')+
   \int \ud[4]{y} \ud[4]{y'} G_{ac}^{(0)}(x,y) [-i\Sigma^{cd}(y,y')]   
G_{db}(y',z),
\end{equation}
where $G^{(0)}_{ab}(x,x')$ are the propagators of the free theory and  
$G_{ab}(x,x')$ are the propagators of the full interacting theory.  
Notice that Eq.~\eqref{SelfEnergyGeneral} is matrix equation relating  
the four components of the self-energy with the four components of the  
propagator. Therefore there is no diagonal relation between $G_{11} 
(x,x')$ and $\Sigma^{11}(y,y')$  as in the vacuum case.
The $ab$ component of the self-energy can be computed, similarly
to the vacuum case, as the sum of all one-particle irreducible
diagrams with  amputated external legs that begin and end with
type $a$ and type $b$ vertices, respectively. CTP Feynman rules are  
completed with the prescription of adding one minus sign for every  
type 2 vertex.

A particularly useful combination is the  retarded self-energy,
defined as $\Sigma_\mathrm R(x,x') = \Sigma^{11}(x,x') +
\Sigma^{12}(x,x')$. It  is related to the retarded propagator through
\begin{equation} \label{SelfEnergyGeneralRetarded}
   \GR(x,x') = \GR^{(0)}(x,x')+
   \int \ud[4]{y} \ud[4]{y'} \GR^{(0)}(x,y) [-i\SigmaR(y,y')]   
\GR(y',z),
\end{equation}
This equation can be regarded as a consequence of the causality  
properties of the retarded propagator. A similar relation holds  
between the
advanced propagator $G_\mathrm A(x,x')$ and the
advanced self-energy $\Sigma_\mathrm A(x,x') = \Sigma^{11}(x,x') +  
\Sigma^{21}(x,x')$. Another useful combination is the Hadamard self-energy, which is defined as $\Sigma^{(1)}(x,x') = \Sigma^{11}(x,x') +  
\Sigma^{22}(x,x')$ [or equivalently as $\Sigma^{(1)}(x,x') =-  
\Sigma^{12}(x,x') - \Sigma^{21}(x,x')$]   and which is
related to the Hadamard propagator through \cite{ArteagaThesis}
\begin{equation} \label{GSigmaN}
	G^{(1)}(x,x') = -i \int \ud[4]{y} \ud[4]{y'} \GR(x,y) \SigmaN(y,y')  
\GA(y',x').
\end{equation}
All self-energy combinations can be determined from the knowledge of  
the Hadamard self-energy and the imaginary part of the retarded self-energy. This latter quantity can be obtained from the following  
cutting rule:
\begin{equation}
	\Im \SigmaR(x,y) = \frac{i}{2} \left[\Sigma^{21}(x,y) - \Sigma^{12} 
(x,y) \right].
\end{equation}


So far, all expressions in this appendix refer to arbitrary background  
states $\hat \rho$. For static and homogeneous backgrounds,   
\Eqref{SelfEnergyGeneralRetarded} can be solved for the retarded  
propagator by going to the momentum representation:
\begin{equation}
	\GR(\omega,\vect p) = \frac{-i}{ -\omega^2 + m^2 + \vect p^2+ 
\SigmaR(\omega,\vect p)-p^0 i\epsilon}.
\end{equation}
We have considered that the free propagators of the mode $\vect p$ are  
those corresponding to the action \eqref{SysAction}. Notice that in  
general the self-energy is a separate function of the energy $\omega$  
and the 3-momentum $\vect p$, and not only a function of the scalar  
$p^2$, as in the vacuum case.
The Hadamard function admits the following expression [which can be  
derived from \Eqref{GSigmaN}]:
\begin{equation}
	G^{(1)}(\omega,\vect p) = i |\GR(\omega,\vect p)|^2 \SigmaN(\omega, 
\vect p)  = \frac{i\SigmaN(\omega,\vect p)}
	{ [-\omega^2 + m^2 + \vect p^2+\Re\SigmaR(\omega,\vect p)]^2 + [\Im  
\SigmaR(\omega,\vect p)]^2}.
\end{equation}
From the retarded propagator and the Hadamard function we can derive:
\begin{subequations}\label{PropsSigma}
\begin{align}
   G_\mathrm F(\omega,\vect p)  &= \frac{ -i \left[ -\omega^2 + E_ 
\vect p^2 + \Re \SigmaR(\omega,\vect p)
   \right]
   + i\SigmaN(\omega,\vect p)/2}{\left[ -\omega^2 + m^2 + \vect p^2+ \Re  
\SigmaR(\omega,\vect p)\right]^2 +
   \left[\Im \SigmaR(\omega,\vect p)\right]^2}, \\
   G_\mathrm D(\omega,\vect p)  &= \frac{ i \left[ -\omega^2 + E_ 
\vect p^2 + \Re \SigmaR(\omega,\vect p)
   \right]
   + i\SigmaN(\omega,\vect p)/2}{\left[ -\omega^2 + m^2 + \vect p^2+ \Re  
\SigmaR(\omega,\vect p)\right]^2 +
   \left[\Im \SigmaR(\omega,\vect p)\right]^2}, \\
   G_-(\omega,\vect p) &= \frac{i\SigmaN(\omega,\vect p)/2 + \Im  
\SigmaR(\omega,\vect p)}{\left[ -\omega^2 + m^2 + \vect p^2+ \Re  
\SigmaR(\omega,\vect p)\right]^2
   +
   \left[\Im \SigmaR(\omega,\vect p)\right]^2}, \\
   G_+(\omega,\vect p) &= \frac{i \SigmaN(\omega,\vect p)/2 - \Im  
\SigmaR(\omega,\vect p)}{\left[ -\omega^2 + m^2 + \vect p^2+ \Re  
\SigmaR(\omega,\vect p)\right]^2
   +
   \left[\Im \SigmaR(\omega,\vect p)\right]^2}.
\end{align}
\end{subequations}
When the field state is not exactly homogeneous, the above expressions  
are still correct up to order $Lp$, where $L$ is the relevant  
inhomogeneity time or length scale.

\section{Linear open quantum systems} \label{app:QBM}

We present those aspects of the theory of linear open quantum systems  
relevant in this paper, focusing on the propagators of the Brownian  
particle. For a more complete presentation check Refs.~\cite 
{Davies,BreuerPetruccione,GardinerZoller,CaldeiraLeggett83a,CalzettaRouraVerdaguer03,Weiss}.

We shall consider a quantum Brownian motion (QBM) model: an open  
quantum system composed of a harmonic
oscillator $q(t)$, which will be the subsystem under study,
linearly coupled to a free massless field $\varphi(t,x)$, which will
act as environment or reservoir. The action for the full system
can be decomposed in the action of the harmonic oscillator, the action
of the scalar field and the interaction term as
\begin{subequations}
\begin{align}
   S_\mathrm{sys}[q] & = \int \ud{t} \left[ \fud \dot q^2 - \fud      
\Omega^2 q^2 \right], \\
   S_\mathrm{env}[\varphi] & = \int \ud{t} \ud{x}   
\left[ \fud(\partial_t
   \varphi)^2-\fud(\partial_x \varphi)^2 \right], \\
   S_{\text{int}}[q,\varphi] & = g \int \ud{t} \ud{x} \delta(x)  \dot q
   \varphi,  \label{SInt}
\end{align}
\end{subequations}
with  $\Omega$ being the frequency of the
harmonic oscillator  and $g$ being the coupling
constant. The oscillator is taken to have unit mass. A counterterm  
action including a frequency shift could also be considered.

We use a one-dimensional free field as the environment, following the  
treatment of Ref.~\cite{UnruhZurek89}. This is equivalent to the  
alternative representation
in which the environment is modelled by a
large ensemble of  harmonic oscillators  \cite{CaldeiraLeggett83b}. This
equivalence can be seen  performing a mode decomposition in the
interaction term \eqref{SInt},
\begin{equation}
   S_{\text{int}}[q,\varphi] = \sqrt{L} \int \ud t \frac{\vd p}{2\pi}  
\glin \dot q
   \varphi_p,
\end{equation}
where $\varphi_p(t)$ is proportional to the spatial Fourier transform  
of the
scalar field,
\begin{equation*}
   \varphi_p(t) = \frac{1}{\sqrt{L}} \int \ud x \expp{- i p x}  
\varphi(t,x),
\end{equation*}
where $L$ is the length of the real axis (formally infinite).

\index{Distribution of frequencies}
\index{Distribution of frequencies!ohmic}
The standard model, also called ohmic model, can be generalised by  
replacing the delta interaction of
equation \eqref{SInt} by a function $f(x)$. In this case the  
interaction term is,
\begin{subequations}
\begin{equation}
   S_\text{int}[q,\varphi] = \int \ud t \ud x f(x)g \dot q(t)  
\varphi(t,x).
\end{equation}
or equivalently in the Fourier space,
\begin{equation}
   S_\text{int}[q,\varphi] = \sqrt{L} \int \ud t  \frac{\ud p}{2\pi}  
\tilde f(-p) g \dot{q}(t)
   \varphi_p(t).
\end{equation}
\end{subequations}

The real even function $\mathcal I(p)= \tilde f(p) \tilde f(-p)$ is  
called the distribution of frequencies.\footnote{In the literature the  
distribution of frequencies is frequently defined as $\omega\mathcal  
I(\omega)$.} The product $\glin^2 \mathcal I(\omega)$ characterises  
the properties of the coupling with the environment at a given energy $ 
\omega$. The QBM model this way generalised encompasses the entire  
class of linearly coupled environments.

When the system and the environment are
initially uncorrelated, {\em i.e.}, when the initial density
matrix
factorises ---$\hat{\rho}(t_\mathrm i)=\hat{\rho}_\mathrm{s}(t_\mathrm  
i)\otimes \hat{\rho}
_\mathrm{e}(t_\mathrm i)$, where $\hat{\rho}_\text{s}(t_\mathrm i)$
and $\hat{\rho}_\text{e}(t_\mathrm i)$ mean, respectively, the density
matrix operators of the system and the environment at the initial
time---, the evolution for the reduced density matrix can be written as
\begin{equation}
   \rho _\text{s}(q_\mathrm f,q_\mathrm f^{\prime },t_\mathrm f)=\int  
\ud{q_\mathrm i}\ud{q_\mathrm i^{\prime
   }}J(q_\mathrm f,q_\mathrm f^{\prime },t_\mathrm f;q_\mathrm i,q_ 
\mathrm i^{\prime },t_\mathrm i)\rho
   _\text{s}(q_\mathrm i,q_\mathrm i^{\prime },t_\mathrm i)\text{,}
\end{equation}
where the propagator $J$ is defined in a path integral
representation by
\begin{equation}
   J(q_\mathrm f,q_\mathrm f^{\prime },t_\mathrm f;q_\mathrm i,q_ 
\mathrm i^{\prime
   },t_\mathrm i)=\int\limits_{q(t_\mathrm i)=q_\mathrm i}^{q(t_ 
\mathrm f)=q_\mathrm f}{\cal D}
   q\int\limits_{q^{\prime }(t_\mathrm i)=q_\mathrm  
i^{\prime }}^{q^{\prime
   }(t_\mathrm f)=q_\mathrm f^{\prime }}{\cal D}q^{\prime }  
\expp{i(S[q]-S[q^{\prime
   }]+S_{\mathrm{IF}}[q,q^{\prime }]) }\text{,}
\end{equation}
with $S_{\mathrm{IF}}[q,q^{\prime }]$ being the influence action,
which is related to the the influence functional introduced by Feynman  
and
Vernon \cite{FeynmanVernon63,FeynmanQMPI} through $F[q,q^{\prime
}]=\expp {iS_{\mathrm{IF}}[q,q^{\prime }]}$. In turn, the
influence functional can be expressed in the following way:
\begin{equation} \label{FInfl}
\begin{split}
    F[q,q^{\prime }]  =  \iint & \D \varphi \, \D \varphi'
    \rho_\mathrm{e}([\varphi_\mathrm i],[\varphi'_\mathrm i],t_ 
\mathrm i)  \exp{\left[ i \left(
    S[\varphi] - S[\varphi'] + S_{\text{int}}[q,\varphi] -
    S_{\text{int}}[q',\varphi'] \right) \right]}.
\end{split}
\end{equation}
The path integral has the boundary conditions
$\varphi(x,t_\mathrm i)=\varphi_\mathrm i(x)$, $\varphi'(x,t_\mathrm
i)=\varphi'_\mathrm i(x)$, $\varphi(x,t_\mathrm  f)=\varphi'(x,t_\mathrm
f)=\varphi_\mathrm f(x)$; there is also an implicit sum over initial
and final states, $\varphi_\mathrm i(x)$, $\varphi_\mathrm i'(x)$ and
$\varphi_\mathrm f(x)$.

Considering a factorized initial state is a rather unphysical hypothesis that leads to surprising results in many circumstances (see for instance Ref.~\cite{HuPazZhang92}). The methods presented in this appendix can be generalized to more natural initial density matrices by the use of the so-called preparation functions \cite{Weiss,GrabertEtAl88}. However the preparation function method does not completely solve all the problems because it is based in a sudden change of the density matrix. A more physical approach involves a continuous preparation of the system \cite{AnglinPazZurek96}. In any case, these techniques are increasingly more involved, and we shall be only interested in studying the dynamics much after the typical decoherence time. In this case the system and environment have had enough time to interact and become entangled, and the precise form of the initial state becomes unimportant.

When the initial density matrix of the environment $\rho
_\mathrm{e}([\varphi_\mathrm i],[\varphi_\mathrm i'],t_\mathrm i)$ is
Gaussian, the path integrals can be exactly performed and one
obtains \cite{FeynmanVernon63,CaldeiraLeggett83a,RouraThesis}:
\begin{equation}
\begin{split}
   S_{\mathrm{IF}}[q,q^{\prime}]
   =&- 2\int_{t_\mathrm i}^{t_\mathrm f} \ud{t} \int_{t_\mathrm  
i}^{t} \ud{t^{\prime
   }}\Dot \Delta (t) \mathcal D(t,t^{\prime })\dot Q(t^{\prime })  +  
\frac{i}{2}\int_{t_\mathrm i}^{t_\mathrm f}\ud{t}
   \int_{t_\mathrm i}^{t_\mathrm f} \ud{t^{\prime }}\Dot \Delta (t)  
\mathcal N(t,t^{\prime })\Dot \Delta
   (t^{\prime })\text{,}
\end{split}
\end{equation}
where $\Delta(t) =  q(t)- q'(t)$ and $Q(t) = [ q(t)+
q'(t)]/2$.
\index{Dissipation kernel}
\index{Noise kernel}

The kernels can be computed as
\begin{subequations} \label{mathcals}
\begin{align}
   \mathcal D(t,t')&=   \frac{i \glin^2}{2} \int \frac{\vd p}{2\pi}  
\mathcal I(p) \Tr \left( \hat\rho\,[\hat \varphi_{\mathrm I(-p)} (t),  
\hat
   \varphi_{\mathrm Ip} (t')]  \right)
    , \label{mathcalD} \\
   \mathcal N(t,t')&= \frac{ \glin^2}{2} \int \frac{\vd p}{2\pi}  
\mathcal I(p) \Tr \left( \hat\rho \, \{\hat \varphi_{\mathrm I(-p)}  
(t), \hat
   \varphi_{\mathrm Ip} (t')\}  \right), \label{mathcalN}
\end{align}
\end{subequations}
where $\hat \varphi_\mathrm I (x,t)$ is the field operator in the
interaction picture and $\hat \varphi_{\mathrm Ip} (t)$ is the $p$-mode of the same field operator.\index{Dissipation kernel}
\index{Noise kernel}
By integration by parts, the influence action can also be
expressed as
\begin{equation} \label{S_IF2}
\begin{split}
   S_{\mathrm{IF}}[q,q^{\prime}]=
   & \int_{t_\mathrm i}^{t_\mathrm f} \ud{t} \int_{t_\mathrm i}^{t_ 
\mathrm f} \ud{t'} \Delta(t) H(t,t') Q(t')   + \frac{i}{2} \int_{t_ 
\mathrm i}^{t_\mathrm f} \ud{t} \int_{t_\mathrm i}^{t_\mathrm f}  
\ud{t'} \Delta(t) N(t,t')
   \Delta(t'),
\end{split}
\end{equation}
or as
\begin{equation} \label{S_IF3}
\begin{split}
   S_{\mathrm{IF}}[q,q^{\prime}]
   =&- 2\int_{t_\mathrm i}^{t_\mathrm f} \ud{t} \int_{t_\mathrm  
i}^{t} \ud{t^{\prime
   }}\Delta (t) D(t,t^{\prime }) Q(t^{\prime }) +  \int_{t_i}^{t_f}  
\ud{t} \delta\Omega^2 \Delta(t)Q(t) \\ & + \frac{i}{2}\int_{t_ 
\mathrm i}^{t_\mathrm f}\ud{t}
   \int_{t_\mathrm i}^{t_\mathrm f} \ud{t^{\prime }} \Delta (t)  
N(t,t^{\prime }) \Delta
   (t^{\prime })\text{,}
\end{split}
\end{equation}
where the different kernels are defined as
\begin{subequations} \label{HDN}
\begin{align}
	H(t,t') &=-2 \derp{}{t} \derp{}{t'} [\theta(t-t') \mathcal D(t,t')]  
\label{kernelH1} \\  &= -2\theta(t-t') D(t-t') + \delta \tilde\Omega^2  
\delta(t-t') \label{kernelH},
	\\
	D(t,t') &=   \derp{}{t} \derp{}{t'}  \mathcal D(t,t')   
\label{kernelD},\\
	N(t,t') &=   \derp{}{t} \derp{}{t'} \mathcal N(t,t') \label{kernelN}.
\end{align}
The kernels $D(t,t')$ and $N(t,t')$ are called respectively
dissipation and noise kernels. The frequency shift $\delta\tilde\Omega^2$  
is a formally divergent quantity given by
\begin{equation}\label{FreqShift}
	\delta \tilde\Omega^2 = 2 \lim_{t\to t'} \derp{\mathcal D(t,t')}{t} .
\end{equation}
\end{subequations}

The dissipation and noise kernels can be computed following Eqs.~\eqref{mathcals} and \eqref{HDN}. The value of the dissipation kernel  
in the frequency space is  \cite{ArteagaThesis}:
\begin{equation} \label{DisQBM}
   D(\omega) = \frac{i \omega  \glin^2}{2} \mathcal I(\omega).
\end{equation}
The dissipation kernel is closely related to the kernel $H(\omega)$  
[see Eqs.~\eqref{kernelH1} and \eqref{kernelH}], which is also state-independent and given by:
\begin{equation}
	H(\omega) = \glin^2 \int \frac{\vd p}{2\pi} \frac{\omega \mathcal  
I(\omega)}{\omega-\omega'+i\epsilon} + \delta \tilde\Omega^2.
\end{equation}
In contrast, the noise kernel \eqref{kernelN} is state-dependent. For  
a general Gaussian stationary environments, characterised by the  
occupation numbers $n(p) = \Tr{ \big( \hat\rho_\text e \hat a^\dag_p  
\hat a_p \big)}$, the noise kernel in Fourier space is given by
\begin{equation} \label{NoiseQBM}
	N(\omega)=  {= \glin^2|\omega|}\mathcal I(\omega)\left[\frac12 + n(| 
\omega|)\right].
\end{equation}
The dissipation and noise kernels are related through:
\begin{equation}
	N(\omega) = - i \sign(\omega) \left[\frac12+n(|\omega|)\right] D(\omega).
\end{equation}
For the particular case of an environment in thermal equilibrium at a  
temperature $T$ the occupation numbers are given by $n(|\omega|) = 1/ 
(\expp{|\omega|/T}-1)$ and the above equation becomes the 
{fluctuation-dissipation theorem}:
\begin{equation}\label{FluctDisTh}
   N(\omega) = - i \sign(\omega) \coth  \left(\frac{|\omega|}{2
   T}\right)
   D(\omega).
\end{equation}

By considering an arbitrary distribution of frequencies $\mathcal  
I(\omega)$ and an arbitrary Gaussian state for the environment $\hat 
\rho_\text{e}$ the dissipation and noise kernels may adopt almost any  
value. In the rest of the appendix we shall try to express all results  
in terms of the dissipation and noise kernels.  To this end, it will  
prove useful to reexpress \Eqref{kernelH} in Fourier space:
\begin{subequations}
\begin{equation}
    H(\omega) =
    - 2 \int \frac{\ud {\omega'}}{2\pi}
    \frac{ i D(\omega')}
    {\omega-\omega'+i\epsilon} + \delta\tilde\Omega^2.
\end{equation}
The kernel $H(\omega)$ can be decomposed in its real and imaginary  
parts as:
\begin{align}
    H_\mathrm R(\omega) &= \Re H(\omega)= -2\PV \int \frac{\ud  
{\omega'}}{2\pi} \frac{
    i D(\omega')}{\omega-\omega'}+ \delta\tilde\Omega^2, \label{RealImOm}  
\\
    H_\mathrm I(\omega) &= \Im H(\omega) =  i D(\omega).
\end{align}
\end{subequations}
We have used the property $1/(x+i\epsilon) = \PV(1/x) - i \pi
\delta(x)$. Notice that $H(-\omega)=H^*(\omega)=
H_\mathrm R(\omega)-iH_\mathrm I(\omega)$. The real and imaginary  
parts of the kernel in frequency space correspond, respectively, to  
the even and odd parts in time space. \index{Kramers-Kronig relation}  
Notice also the Kramers-Kronig relation between the real and imaginary  
parts of the kernel $H(\omega)$:
\begin{equation}
 H_\mathrm R(\omega) = -2\PV \int \frac{\ud {\omega'}}{2\pi} \frac{
    H_\mathrm I(\omega')}{\omega-\omega'}+ \delta\tilde\Omega^2.
\end{equation}
The frequency shift can be always absorbed in $\Omega$. From now on, and in the main body of the paper, we will assume that such absorption has been carried out.

For a Gaussian environment and asymptotic initial boundary
conditions the generating functional can be expressed as\footnote{In  
the general case there would be a prefactor taking into account the  
initial conditions for the system. However, since the system has a  
dissipative dynamics and the initial conditions are given in the  
remote past, initial conditions for the system turn out to be  
completely irrelevant.} \cite{CalzettaRouraVerdaguer03}
\begin{equation}
\begin{split}
   Z[j_1,j_2] =  & \
    \expp{  \fud \int \ud{t_1} \ud{t_2} \ud{t_3}
   \ud{t_4}
    j_\Delta(t_1)
   \Gret(t_1,t_2)N(t_2,t_3)j_\Delta(t_4)\Gret(t_4,t_3) }
   \expp{  -\int \ud{t_1} \ud{t_2} j_\Delta(t_1) \Gret(t_1,t_2)
   j_\Sigma(t_2)},
\end{split}
\end{equation}
where $j_\Sigma(t) = [j_1(t)+j_2(t)]/2$, $j_\Delta(t) =
j_1(t) - j_2(t)$ and $\Gret(t,t')$ is the retarded propagator of the
kernel
\begin{equation} \label{KernelL}
   L(t,t') = \left( \dert[2]{}{t} + \Omega^2 \right)
   \delta(t-t') + H(t,t'),
\end{equation}
\ie, is the kernel which verifies
\begin{equation} \label{RetEqDif}
   \int \ud s \Gret(t,s) L(s,t') = - i \delta(t-t'),
   \qquad \Gret(t,t') = 0 \quad \text{if } t<t'.
\end{equation}
Explicit expressions are most easily obtained in the Fourier space, in  
which the retarded propagator adopts the form:
\begin{equation}\label{GretFourier}
	\GR(\omega) = \frac{-i}{L(\omega)} = \frac{-i}{-\omega^2 +  
\Omega^2  + H(\omega)}.
\end{equation}
It can be checked that $\GR(t,t')$, besides being the retarded  
propagator of the kernel $L(t,t')$, in the sense of \Eqref{RetEqDif},  
is also the retarded propagator of the quantum mechanical system, in  
the sense of \Eqref{RetardedProp}.

\index{Propagator!Feynman}
\index{Propagator!Dyson}
\index{Propagator!Whightman}
\index{Propagator!in the QBM models}

Differentiating the CTP generating functional, according to Eqs.~\eqref{CTPPropSet}, we  obtain the
different correlation functions in terms of the noise and dissipation  
kernels:
\begin{subequations} \label{GOtherFourier}
\begin{align}
   G_\mathrm F(\omega)
   &= \frac{- i \left[ - \omega^2 + \Omega^2  + H_\mathrm R(\omega)  
\right]
   + N(\omega)}{\left[ - \omega^2 + \Omega^2 + H_\mathrm R(\omega) 
\right]^2 +
   \left[H_\mathrm I(\omega)\right]^2}, \\
   G_\mathrm D(\omega)  &= \frac{ i \left[ - \omega^2 + \Omega^2  +  
H_\mathrm R(\omega)
   \right]
   + N(\omega)}{\left[ - \omega^2 + \Omega^2 + H_\mathrm R(\omega) 
\right]^2 +
   \left[H_\mathrm I(\omega)\right]^2}, \\
   G_-(\omega) &= \frac{ N(\omega) + H_\mathrm I(\omega)}{\left[ -  
\omega^2 + \Omega^2 + H_\mathrm R(\omega)\right]^2
   +
   \left[H_\mathrm I(\omega)\right]^2}, \\
   G_+(\omega) &= \frac{ N(\omega) - H_\mathrm I(\omega)}{\left[ -  
\omega^2 + \Omega^2 + H_\mathrm R(\omega)\right]^2
   +
   \left[H_\mathrm I(\omega)\right]^2}.
\end{align}
\end{subequations}
The generating functional can be alternatively expressed in terms of  
these correlation functions as
\begin{equation} \label{ZQBMDirectBasis}
\begin{split}
   Z[j_1,j_2] =  N \expp{-\frac{1}{2} \int \ud{t}\ud{t'}j^a(t) G_{ab} 
(t,t') j^b(t')}.
\end{split}
\end{equation}

If desired, the dynamics of the Brownian oscillator can be analyzed in  
terms of a Langevin equation,
\begin{equation} \label{Langevin}
	\ddot q(t) + \Omega^2 q(t) + \int_{t_\text{i}}^\infty \ud{t'}  
H(t,t') q(t') = \xi(t),
\end{equation}
where $\xi(t)$ is a stochastic Gaussian field defined by the  
correlation functions
\begin{equation}
	\langle \xi(t) \rangle_\xi = 0, \qquad \langle \xi(t) \xi(t') \rangle_ 
\xi = N(t,t'),
\end{equation}
with $\langle \cdots \rangle_\xi$ meaning stochastic average. The  
stochastic correlation functions derived from the Langevin equation  
correspond to a subset of the quantum correlation functions  
\cite{CalzettaRouraVerdaguer03,GardinerZoller}.

It is also possible to study the master equation for the reduced density matrix of  
the system  $\rho_\text s$, which is given by  
\cite{HuPazZhang92,HalliwellYu96,CalzettaRouraVerdaguer03}:
\begin{equation} \label{master}
\begin{split}
	i \derp{}{t} \rho_\text s(q,q',t) &= \bigg[-\frac12 \left( \derp[2]{} 
{q} - \derp[2]{}{q'} \right) + \frac12 [\Omega^2 + \delta  
\Omega_0^2(t)](q^2-q'^2) \\
	&\quad  - i \Gamma(t)(q-q') \left( \derp{}{q} - \derp{}{q'} \right) -  
i \Gamma(t) h(t) (q-q')^2 \\ &\quad + \Gamma(t)f(t)(q-q')  
\left( \derp{}{q} + \derp{}{q'} \right) \bigg] \rho_\text s(q,q',t).
\end{split}
\end{equation}
where $\delta \Omega_0^2(t)$, $\Gamma(t)$, $h(t)$ and $f(t)$ are a  
frequency shift, a dissipative factor and two dispersive factors,  
respectively, which in the weak coupling limit are given by  
\cite{HuPazZhang92}:
\begin{subequations} \label{mastercoeff}
\begin{align}
	\delta \Omega^2 (t)  &= -2 \int_{t_\text i}^t D(s,t_\text i) \cos{\Omega  
(s-t_{\text i})}\,  \vd s,\\
	\Gamma(t) &= \frac{1}{\Omega}  \int_{t_\text i}^t D(s, t_i)  
\sin{\Omega (s-t_{\text i})}\, \vd s,\\
	\Gamma(t)h(t) &= \int_{t_\text i}^t N(s,t_i) \cos{\Omega (s- 
t_{\text i})} \, \vd s,	\\
	\Gamma(t)f(t) &= \frac{1}{\Omega} \int_{t_\text i}^t N(t_i,s)  
\sin{\Omega (s-t_{\text i})} \,\vd s.
\end{align}
\end{subequations}
The corresponding expressions valid for arbitrary couplings can be  
found in Refs.~\cite{HuPazZhang92,CalzettaRouraVerdaguer03}. Recall  
that the time $t_i$ is the time at which the density matrix is assumed  
to factorize.

Alternatively, it is also possible to introduce the reduced Wigner  
function,
\begin{equation}
	W_{\text s}(q,p,t)=\frac1{2\pi} \int d\Delta \expp{i \Delta p}  
\rho_{\text s}(q-\Delta/2,q+\Delta/2,t),
\end{equation}
in terms of which the master equation adopts a Fokker-Planck form:
\begin{equation}
	\derp{W_{\text s}}{t} = - p \derp{W_{\text s}}{q} + [\Omega^2 +  
\delta\Omega_{0}^2(t)]q \derp{W_{\text s}}{p} + 2\Gamma(t)  
\derp{pW_{\text s}}{p}+\Gamma(t)h(t) \derp[2]{W_{\text s}}{p}+ 
\Gamma(t)f(t) \derpp{W_{\text s}}{p}{q}.
\end{equation}
The Wigner function has some similarities with a classical  
distribution function, although it cannot be interpreted as a true  
probability density, among other reasons, beacuse it can adopt negative values \cite{GardinerZoller}.

\end{document}